\documentclass[12pt]{article}
\usepackage{amsmath}
\usepackage{bbm}
\usepackage{graphicx}
\usepackage{enumerate}
\usepackage[round]{natbib}
\PassOptionsToPackage{hyphens}{url}\usepackage{hyperref}
\usepackage{url}[hyphens]
\usepackage{bm}
\usepackage{xcolor}
\usepackage{hyperref}
\usepackage{ulem}
\usepackage{verbatim}
\hypersetup{
  colorlinks   = true, 
  urlcolor     = blue, 
  linkcolor    = blue, 
  citecolor   = blue 
}

\newcommand{\blind}{0}
\renewcommand{\hat}{\widehat}
\renewcommand{\tilde}{\widetilde}
\renewcommand{\bar}{\overline}

\begin{document}
\date{}

\def\spacingset#1{\renewcommand{\baselinestretch}%
{#1}\small\normalsize} \spacingset{1}

\if0\blind
{
    \title{Bayesian Unit-level Models for Longitudinal Survey Data under Informative Sampling: An Analysis of Expected Job Loss Using the Household Pulse Survey}
    \author{
Daniel Vedensky\footnote{(\baselineskip=10pt to whom correspondence should be addressed)
Department of Statistics, University of Missouri,
146 Middlebush Hall, Columbia, MO 65211-6100, dvedensky@mail.missouri.edu},
Paul A. Parker\footnote{\baselineskip=10pt 
Department of Statistics, University of California Santa Cruz,
1156 High St, Santa Cruz, CA 95064, paulparker@ucsc.edu},
   and Scott H. Holan\footnote{\baselineskip=10pt Department of Statistics, University of Missouri,
146 Middlebush Hall, Columbia, MO 65211-6100, holans@missouri.edu}\,\footnote{\baselineskip=10pt U.S. Census Bureau, 4600 Silver Hill Road, Washington, D.C. 20233-9100, scott.holan@census.gov},
\\
}
  \maketitle
}\fi

\if1\blind
{
  \bigskip
  \bigskip
  \bigskip
  \begin{center}
    {\LARGE\bf Bayesian Unit-level Models for Longitudinal Survey Data under Informative Sampling with Application to the Household Pulse Survey}
  \end{center}
  \medskip
} \fi

\bigskip
\begin{abstract}
The Household Pulse Survey (HPS), recently released by the U.S. Census Bureau, gathers timely information about the societal and economic impacts of coronavirus. The first phase of the survey was quickly launched one month after the beginning of the coronavirus pandemic and ran for 12 weeks. To track the immediate impact of the pandemic, individual respondents during this phase were re-sampled for up to three consecutive weeks. Motivated by expected job loss during the pandemic, using public-use microdata, this work proposes unit-level, model-based estimators that incorporate longitudinal dependence at both the response and domain level. In particular, using a pseudo-likelihood, we consider a Bayesian hierarchical unit-level, model-based approach for both Gaussian and binary response data under informative sampling. To facilitate construction of these model-based estimates, we develop an efficient Gibbs sampler. An empirical simulation study is conducted to compare the proposed approach to models that do not account for unit-level longitudinal correlation. Finally, using public-use HPS micro-data, we provide an analysis of ``expected job loss" that compares both design-based and model-based estimators and demonstrates superior performance for the proposed model-based approaches.\\
\end{abstract}

\noindent%
\vfill

\newpage
\spacingset{1.9}

\newpage
\section{Introduction}
\label{sec:intro}

The U.S. Census Bureau launched the Household Pulse Survey (HPS) as an experimental data product shortly after the outbreak of the COVID-19 pandemic in 2020.
Given the unprecedented effects of the outbreak, which required quick decision-making, the HPS was intended to allow for a short turnaround time between the data collection stage and the data publication stage. 
To track these effects, the survey includes a wide variety of questions ranging from economic impacts (e.g., applications for unemployment insurance and whether the respondent received a stimulus check) to health effects (e.g., symptoms of anxiety).

The first phase of HPS, which we focus on in the present work, ran for 12 weeks from April 23 to July 21, 2020 and followed a longitudinal design, where respondents could be interviewed multiple times in quick succession. 
If a household was sampled in a given week and provided a response, that household was retained in the sampling frame for the following week. 
If the household then provided a second response, it was retained in the sampling frame for a third and final week, after which it was dropped. Adjustments for unit-non-response are incorporated through the survey weights.\footnote{\baselineskip = 10pt \url{https://www2.census.gov/programs-surveys/demo/technical-documentation/hhp/2020_HPS_Background.pdf}}

This survey design leads to a structure where households may appear in the data up to three consecutive times. 
We refer to the resulting within-respondent dependence as ``longitudinal correlation'' to distinguish it from the ``temporal correlation'' that arises from the survey being carried out weekly.

The HPS has proven useful for investigating the initial ramifications of the outbreak. 
The CDC has used HPS to measure the effect of the pandemic on mental health conditions like depression and anxiety \citep{Vahratian_2021}.
The Bureau of Labor Statistics has used the survey to track receipt and usage of government stimulus payments.\footnote{\baselineskip=10pt \url{https://www.bls.gov/opub/btn/volume-9/receipt-and-use-of-stimulus-payments-in-the-time-of-the-covid-19-pandemic.htm}} 
Meanwhile, other researchers have used  HPS to analyze a variety of different outcomes ranging from food insecurity \citep{schanzenbach2020much} to education and economic inequality \citep{Francis_2021}.
However, these studies all treat the data as cross-sectional, thereby disregarding the temporal dependence across weeks and the valuable information provided by the longitudinal study design.

Furthermore, the rich demographic data collected by the HPS leads to a small-domain problem naturally amenable to unit-level modeling.
However, the existing literature for unit-level modeling of longitudinal data tends to address only the case of Gaussian responses, despite the ubiquity of binary-valued responses in survey data (e.g., \citet{guadarrama_2020}).
Current methodology that does allow for binary responses, generally ignores the complex survey design and survey weights (e.g., \citet{Hobza_2016,Hobza_2017}), although \cite{kunihama_2019} consider a mixture model that adjusts the mixture component probabilities using the reported survey weights. Since survey samples are unlikely to be representative, or may exhibit informative sampling, models that ignore the survey design and weights are prone to produce bias and imprecise estimates.
Though the literature for fitting unit-level models to complex survey data under informative sampling has seen many recent developments (e.g., see \citet{Parker_2019}), it has thus far focused primarily on cross-sectional data. 
As such, existing unit-level methodology is not directly applicable to longitudinal or time-series data such as the HPS.

Motivated by the HPS and the impacts of expected job loss during the initial phase of the COVID-19 pandemic, we propose unit-level modeling methodology that accounts for both the longitudinal structure and complex survey design. Specifically, we propose models for both Gaussian and binary data, allowing subject-matter researchers to answer a broad range of scientific and policy-related questions.

The paper proceeds as follows. Section~\ref{sec:methods} introduces our notation and methodology. 
In Section~\ref{sec:empsim}, we present an empirical simulation study using existing data from the HPS. 
Section~\ref{sec:hps_analysis} presents results from our motivating data analysis using the full HPS phase one dataset to investigate expected job loss during the initial phase of COVID-19 in the US. 
Finally, we present discussion and some concluding remarks in Section~\ref{sec:discussion}.

\section{Methodology}\label{sec:methods}

\subsection{Notation and Background}
Following the notation of \citet{parker_holan_janicki_2022}, we let $\mathcal{U}=\{1,\ldots,N\}$ denote an enumeration of a finite population.
We take $\mathcal{U}$ to be the union of disjoint subpopulations $\mathcal{U}_j$, $j=1,\ldots, k$, each with cardinality $|\mathcal{U}_j|=N_j$,  so that $\sum_{j=1}^k N_j= N$. 
At each time $t=1,\ldots, T$, a sample $S_t$ of size $n_t$ is taken, with a given (known) probability of each unit being sampled denoted $\pi_{it} = P(i\in \mathcal{S}_t)$.
The index set for units sampled at time $t$  in area $j$ is then $\mathcal S_{jt} =  \mathcal{U}_j \cap \mathcal{S}_t$. 
We denote the entire sample by $\mathcal{S}=\bigcup_{t=1}^{T} \mathcal{S}_t$ with $|\mathcal{S}|=\sum_{t=1}^T n_t=n$.

Let $y_{ijt}$ be a characteristic associated with unit $i$ in area $j$ at time $t$. 
Generally, we would like to estimate the population mean (or proportion, in the case that $y_{ijt}$ is binary) of this characteristic in area $j$ at a  time $t$
\[\bar{y}_{\bm\cdot jt}=\frac{1}{N_{j}}\sum_{i\in\mathcal{U}_{j}} y_{ijt}.\] %
Often, a design-based estimator such as the Horvitz-Thompson (HT) estimator \citep{Horvitz1952},
\[\hat{\bar y}_{\bm\cdot jt} = \frac{1}{N_{j}}\sum_{i\in\mathcal{S}_{jt}} w_{it} y_{ijt},\]
is used, where $w_{it}=\frac{1}{\pi_{it}}.$ 
So-called direct estimators, like the HT estimator, use only the observed survey data and ensure desirable properties such as design-unbiasedness and design-consistency.
However, problems arise for fine partitions of the population, where some subpopulations are bound to have small (or zero) sample sizes.
Such subpopulations tend to produce unreliable direct estimates with large standard errors and are  conventionally referred to as ``small areas,'' even if the designation is not strictly geographic.
Moreover, if a small area has zero sample size, no direct estimate can be calculated. 
In these instances, a model-based approach is more suitable so that small areas may ``borrow'' information from other areas through the use of covariates and random effects \citep{rao15}. 

Small area models may be implemented at either the area level or the unit level. 
In the former, direct estimators for each area are used as response values within a model, which links them through area-level covariates.
Due to this reliance on direct estimators, small sample sizes may still be problematic for area-level models.
Therefore, when data are available at the level of responses from individual units, it may be preferable to fit a unit-level model that does not require the use of direct estimates.

In addition, unit-level models have been shown to produce lower standard errors and better interval coverage than area-level models \citep{hid16}. 
Unit-level models also eliminate the need for ad hoc benchmarking when the data are tabulated at one resolution but estimates are required at another.
While area-level models may violate internal consistency in this setting, unit-level models allow for the arbitrary aggregation of units to produce estimates at any resolution.

The basic unit-level model, first introduced by \citet{Battese_1988} was posed as a frequentist, mixed effects model of the form
\[y_{ij} = \bm x_{ij}\bm\beta + v_j + e_{ij},\]
with unknown parameters $\bm \beta$ and area-level random effects $v_j\overset{iid}{\sim} N(0,\sigma^2_v)$ independent of the unit-level random errors $e_{ij}\overset{iid}{\sim} N(0, \sigma^2_e)$.
Unlike area-level models, finite population parameters are not directly parameterized within a unit-level model. Instead, after fitting a unit-level model, predictions can be made for every unit in the population. This effectively results in a synthetic population that may be aggregated to the resolution of any necessary estimates. In a Bayesian setting, this can be done for every MCMC iteration, resulting in a posterior predictive distribution of the estimates (e.g., see \cite{parker_holan_janicki_2022} for details).

However, the basic unit-level model makes the critical assumption that the distribution of sampled units matches the population distribution. 
This rarely holds for complex survey designs.
Often, individual response values are correlated with their probability of being included in the sample, a phenomenon known as informative sampling (IS). %
\citet{Parker_2019} describe the problems that arise when IS is ignored and provide an overview of the most common methods for correcting for it in unit-level models.

One straightforward method for adjusting a unit-level model to an informative sample is via the pseudo-likelihood (PL) \citep{skinner_1989, binder_1983}.
In a PL, the contribution of each sampled unit to the likelihood is exponentially weighted by its survey weight,
\[\text{PL}(\bm y|\theta) = \prod_{i\in \mathcal{S}} f(y_i|\theta)^{w_i}.\]
\citet{savitsky_toth_2016} show that such a likelihood may be included in a Bayesian model and leads to a valid pseudo-posterior distribution when multiplied by a prior density $\pi(\bm\theta)$, 
\[\hat\pi(\bm \theta|\bm y, \tilde{\bm w})\propto \left\{\prod_{i\in \mathcal{S}} f(y_i|\theta)^{w_i}\right\}\pi(\bm \theta),\]
provided the weights are scaled to sum to the sample size at each time, $n_t$,
\[\widetilde{w}_{it}=n_t\frac{w_{it}}{\sum_{\ell\in S_t} w_{\ell t}}.\]
In the cross-sectional setting, \citet{Zhang_2014} fit such a pseudo-likelihood model and used it to produce population estimates, albeit in a frequentist context.

This unit-level modeling approach has been employed for cross-sectional data by \citet{parker_holan_janicki_2022} for categorical data and \citet{Parker_2023} for Gaussian data, who also provide efficient sampling algorithms. \citet{sun_2022} analyze the HPS using this framework, but without considering the longitudinal aspect of the HPS data.

\subsection{Gaussian response model}
For the Gaussian case, we model the within-respondent longitudinal correlation with an AR(1) structure.
This leads to different functional forms for the pseudo-likelihood contribution of first-time responses and follow-up responses since the latter must be conditioned on the previous response. %
To distinguish these two groups in the sample, we let $\mathcal{S}_{t_1}\subset \mathcal{S}_t$ be the index set for respondents who are filling out the survey for the first time at time $t$ and $\mathcal{S}_{t_2} \subset \mathcal{S}_t$ the index set for follow-up respondents---those who are filling out the survey for the second or third time---at time $t.$
The pseudo-likelihood contribution for a unit $i\in S_{t_1}$ is given by
\[y_{it}| \mu_{it},\sigma^2 \propto \text N(y_{it}|\mu_{it}, \sigma^2)^{\widetilde w_{it}},\]%
where we write $N(y|\mu,\sigma^2)$ to denote a Gaussian density function with mean $\mu$ and variance $\sigma^2$ evaluated at $y$.
For $i\in S_{t_2}$ the contribution is given by
\[y_{it}|y_{it-1},\mu_{it},\sigma^2,\rho \propto \text N(y_{it}|\mu_{it} + \rho\widetilde{y}_{i(t-1)}, \sigma^2(1-\rho^2))^{\widetilde w_{it}}.\]%
Here, $\mu_{it}=\bm x_i'\bm\beta + \bm\psi_i'\bm\eta_t$, with $\bm x_i$ denoting the $p-$dimensional vector of known covariates for unit $i$ and $\bm \beta$ the corresponding parameter vector of fixed effects.
The $m-$dimensional vector $\bm\psi_i$  indexes which geographic area unit $i$ belongs to and $\bm \eta_t$ is the corresponding vector of area-level random effects at time $t$.
(Note that $\bm \psi_i$ could be replaced by a more complex set of spatial basis functions.)
Also, $\widetilde{y}_{it}=y_{it}-\mu_{it}$. 
The autoregressive parameter $\rho \in (-1,1)$ reflects the correlation between a current and previous response for a given respondent. %

This pseudo-likelihood is then included in the following hierarchical model, which we refer to as the Gaussian time-dependent unit-level model (G-TULM),
\begin{align*}
\bm y|\bm \beta, \bm \eta, \rho, \sigma^2  &\propto \prod_{t=1}^T\prod_{i\in S_{t_1}} \text{N}(y_{it}|\mu_{it},\sigma^2)^{\widetilde{w}_{it}}\\ 
                                 &\times \prod_{i\in S_{t_2}}\text{N}(y_{it}|\mu_{it} + \rho\widetilde{y}_{i(t-1)},\sigma^2(1-\rho^2))^{\widetilde{w}_{it}}\\
\mu_{it} &= \bm x_i'\bm\beta +\bm\psi_i\bm\eta_t\\
\bm{\beta}    &\sim \text{N}_p(0, \sigma_\beta^2I_p)\\
\bm\eta_1 |\sigma^2_{\eta_1}       &\sim \text{N}_m(\mathbf{0}, \sigma^2_{\eta_1}I_m)\\
\bm\eta_t |\bm\eta_{t-1}, \phi,\sigma^2_{\eta} &\sim \text{N}_m(\phi \bm\eta_{t-1} , \sigma_\eta^2I_m)\\
\rho, \phi    &\overset{ind.}{\sim} \text{Unif}(-1, 1) \\
\sigma^2,\sigma_{\eta_1}^2 , \sigma_\eta^2 &\overset{ind.}{\sim} \text{IG}(a, b)\\
\sigma^2_\beta, a, b &>0.
\end{align*}

The value $\phi$ acts as the lag-1 autoregressive parameter associated with the random effects.
As such, there are two levels of AR(1) structure in the model: one for the temporal random effect governed by $\phi$, the other for the longitudinal correlation governed by $\rho$.%

We estimate the model using a Gibbs sampler, since the full conditional distributions for all of the parameters, except $\rho$, are conjugate. Estimation of $\rho$ leads to a Metropolis within Gibbs step. %
Details for the sampling algorithm and its derivation are provided in the Appendix.

\subsection{Binomial response model}
In the binomial case, we propose the following model, which we call the binomial time-dependent unit-level model (B-TULM),
\begin{align*}
\bm y| \bm\beta, \bm\eta &\propto \prod_{t=1}^T\prod_{i\in \mathcal{S}_t} \text{Binomial}(y_{it} | n_{it}, \xi_{it})^{\widetilde{w}_{it}}\\
\text{logit}(\xi_{it}) &= \bm x_i'\bm\beta+\bm\psi_i\bm\eta_t\\
\bm \beta &\sim \text{N}_p(\bm 0, \sigma^2_\beta I_p)\\
\bm \eta_1 |\sigma^2_{\eta_1} &\sim \text{N}_m(\bm 0, \sigma^2_{\eta_1} I_m)\\
\bm \eta_t |\bm\eta_{t-1},\phi,\sigma^2_{\eta} &\sim \text{N}_m(\phi\bm\eta_{t-1}, \sigma^2_{\eta} I_m), \qquad \text{ for } t=2,\ldots,T\\
\phi &\sim \text{Unif}(-1,1)\\
 \sigma^2_{\eta_1}, \sigma^2_\eta, &\overset{ind.}{\sim} IG(a,b)\\
\sigma_\beta, a, b &> 0,
\end{align*}
where we write Binomial$(y|n,\xi)$ to denote a binomial probability mass function with size $n$ and probability of success $\xi$ evaluated at $y$. Note that for our motivating application, the data is binary, and thus $n_{it}=1$ for all $i$ and $t$.

Unlike the Gaussian case, we cannot easily include an AR(1) structure in the likelihood because the binomial PL does not have an observation error component.
As a result, there is no autocorrelation parameter equivalent to $\rho$ from the previous model.
Instead, to account for the within-respondent correlation, we add a synthetic, categorical covariate to the design matrix. 
This covariate has three levels indexing whether the respondent (1) was sampled and responded ``yes" to the question at the previous time point, (2) was sampled and responded ``no" at the previous time point, or (3) was not sampled in the survey at the previous time point.

We use the P\'olya-Gamma data-augmentation scheme for binomial likelihoods proposed by \citet{polson_scott_2013} to implement a fully Gibbs sampler.
\citet{polson_scott_2013} define a random variable $X$ to have P\'olya-Gamma distribution with parameters $b>0$ and $c\in\mathbbm{R}$, denoted $X\sim \text{PG}(b,c)$, if
\[X\overset{d}{=}\frac{1}{2\pi^2}\sum_{k=1}^\infty\frac{g_k}{(k-1/2)^2+c^2/(4\pi^2)},\]
with $g_k\overset{ind}{\sim} \text{Gamma}(b,1).$ 
They provide an efficient sampling algorithm for P\'olya-Gamma random variables and prove the following integral identity for a binomial likelihood parameterized by the log-odds $\lambda=\text{logit}(\xi_{it})=\bm x_i'\bm\beta +\bm\psi_i'\bm\eta_t$,
\[\frac{(e^\lambda)^a}{(1+e^\lambda)^b}=2^{-b}e^{\kappa\lambda}\int_0^\infty e^{-\omega\lambda^2/2}p(\omega)d\omega,\]
where $\kappa=a-b/2$ and $\omega\sim \text{PG}(b,0)$.
The right-hand side can be viewed as the conditional distribution of $\lambda$  given $\omega$, in which case the integrand is proportional to a Gaussian density.
On the other hand, viewing the expression as the conditional distribution of $\omega$ given $\lambda$ implies a P\'olya-Gamma distribution. 
This suggests a simple Gibbs sampling scheme, alternating draws of Gaussian and P\'olya-Gamma random variables.

Meanwhile, the left-hand side of the integral identity is readily adaptable to our binomial pseudo-likelihood. 
In the B-TULM model, we have
\[\xi_{it}=\frac{\exp\left\{\bm x_i'\bm\beta+\bm \psi_i\bm \eta_t\right\}}{1+\exp\left\{\bm x_i'\bm\beta+\bm \psi_i\bm \eta_t\right\}},\]
which allows the contribution of unit $i$ to the binomial pseudo-likelihood to be written
\begin{align*}
p(y_{it}|n_{it},\xi_{it}) &\propto [\xi_{it}^{y_{it}}(1-\xi_{it})^{n_{it}-y_{it}}]^{\widetilde{w}_{it}} \\
                             &= \frac{(\exp\{\bm x_{i}'\bm\beta+\bm\psi_i\bm\eta_t\})^{y_{it}\widetilde w_{it}}}
                                            {(1+\exp\{\bm x_{i}'\bm\beta+\bm\psi_i\bm\eta_t\})^{n_{it}\widetilde{w}_{it}}}.
\end{align*}
This is simply the left-hand side of the integral identity with $a=y_{it}\widetilde w_{it}$ and $b=n_{it}\widetilde{w}_{it}.$

Introducing P\'olya-Gamma latent variables into the hierarchical model then leads to full conditionals of standard form that can be sampled from directly.
Further details of the sampling algorithm are provided in the Appendix.
\section{Empirical Simulation Studies}\label{sec:empsim}

To assess the performance of our proposed methodology, we conduct an empirical simulation study for both a continuous and a binary response. 
We use public-use microdata files associated with the first 12 weeks of the HPS, available on the U.S.\ Census Bureau website.\footnote{\url{https://www2.census.gov/programs-surveys/demo/datasets/hhp/2020/}} We take our population to be respondents from the contiguous United States, including the District of Columbia. 
In both simulations, we remove respondents who did not provide an answer to the response value under consideration. %

In the Gaussian case we take the response variable as the reported amount for ``household money spent in last 7 days on food to be prepared and eaten at home". These responses take on integer values between $1$ and $900$ and are right-skewed.
Exploratory analysis indicates that application of a Box-Cox transformation leads to an approximately Gaussian distribution.\footnote{\baselineskip=10pt The Box-Cox transformation was conducted using the \texttt{forecast} package, with the parameter $\lambda=-0.0863$ selected automatically \citep{forecast}.}
After removing item non-responses, we are left with 729,219 values.
For the binary response, we appeal to our motivating application and consider the 819,509 responses provided to the question of whether a respondent ``expected household job loss in the next four weeks due to the coronavirus pandemic.''

Both simulations are run for 100 samples.
To preserve the natural pattern of follow-up response rates, we do not sample each week separately.
Instead, we take a single sample from the population without replacement.
If a unit is sampled, all of its provided weekly responses are included in the sample.
In order to make the sample informative, we take a probability proportional to size sample using the Poisson method of \citet{brewer_early_hanif_1984}. 
 We set the expected sample size to be 2\% of the constructed population and sample each unit $i$ in proportion to the size variable
\[s_i=\exp(.1\overline{y}_i+.2w^*_i),\]
where $w^*_i$ is the unscaled original survey weight associated with the  first response of unit $i$ and $\overline{y}_i$ is the average of the responses for unit $i$.

We then fit the model, including gender as well as linear and quadratic terms for age as covariates. 
As noted previously, the binary response model includes an additional synthetic categorical covariate with three levels indicating the type of response recorded for the previous week. 
For the Gaussian response, the Gibbs sampler is run for 2,000 iterations with the first 500 of these discarded as burn-in. 
For the binary response, the sampler is run for 8,000 iterations with the first 1,000 discarded as burn-in. 
We assessed convergence through visual inspection of the trace plots of the sample chains, for a randomly selected subset of parameters, with no lack of convergence detected.%

At each iteration $k$ of the sampler, we generate posterior predictions for each area at each time using population covariate values.
In the Gaussian case we generate
\[\widehat{y}^{(k)}_{it}|\cdot \sim N(\bm x_i'{\bm\beta}^{(k)} + \bm\psi_i'{\bm\eta_t}^{(k)},(\sigma^2)^{(k)})\]
for all units $i$ and times $t$ in the population. 
In the binary case, the procedure is identical but with posterior predictive distribution
\[\widehat y^{(k)}_{it}|\cdot \sim \text{Bernoulli}(\text{logit}^{-1}(\bm x_i'{\bm\beta} +\bm\psi_i'{\bm\eta}_t)).\]

We then aggregate these estimates to obtain a mean for each combination of a state $j$ and a week $t$,
\[\hat{\bar{y}}^{(k)}_{jt} = \frac{1}{|\mathcal{U}_j|}\sum_{i\in \mathcal{U}_j} \hat{y}_{it}^{(k)}.\]
Since this is performed for all iterations of the sampler, we are able to get a posterior  distribution of the domain-level estimates from which we can calculate point estimates and credible intervals.

As a baseline for comparison, we calculate direct estimates for each area at each week and fit a cross-sectional unit-level model separately for each week. 
In the Gaussian case we compare to the pseudo-likelihood model used by \citet{Parker_2023}, which is a Bayesian extension of the basic unit-level model proposed by \citet{Battese_1988}
\begin{align*}
\bm y | \bm\beta,\bm\eta &\propto \prod_{i \in S_t} \text{N}(y_i|\mu_i,\sigma^2)^{\widetilde w_{i}}\\
\mu_i &= \bm x_i'\bm\beta +\bm\psi_i'\bm\eta\\
\bm\eta|\sigma^2_\eta &\sim N_m(\bm 0_m, \sigma^2_\eta\bm I_m)\\
\bm\beta &\sim N_p(\bm 0_p, \sigma^2_\beta\bm I_p)\\
\sigma^2_\eta &\sim \text{IG}(a,b)\\
\sigma^2_\beta, a,b &> 0.
\end{align*}
We refer to this as the Gaussian baseline unit-level model (G-BULM).%

In the binary case, we use the binary unit-level model of \citet{parker_holan_janicki_2022}, which we call the binary baseline unit-level model (B-BULM)
\begin{align*}
\bm y | \bm\beta,\bm\eta &\propto \prod_{i \in S_t} \text{Binomial}(y_i|n_i,\xi_i)^{\widetilde w_{i}}\\
\text{logit}(\xi_i) &= \bm x_i'\bm\beta +\bm\psi_i'\bm\eta\\
\bm\eta|\sigma^2_\eta &\sim N_m(\bm 0_m, \sigma^2_\eta\bm I_m)\\
\bm\beta &\sim N_p(\bm 0_p, \sigma^2_\beta\bm I_p)\\
\sigma^2_\eta &\sim \text{IG}(a,b)\\
\sigma^2_\beta, a,b &> 0.
\end{align*}

In fitting all four models, we set $a=b=1$ and $\sigma^2_\beta=10^4$ in order to have vague prior distributions on $\bm \beta$, $\sigma^2_{\eta_1}$, and $\sigma^2_\eta$. 
For each sub-sample and each population quantity, we use the posterior mean as a point estimate, and construct a 95\% credible interval to quantify uncertainty around the point estimate. We evaluate the quality of our point estimates in terms of mean squared error (MSE), as it strikes a balance between bias and variance. We evaluate the quality of our credible intervals using the interval score  \citep{Gneiting_2007}, which, for a given $(1-\alpha)\times 100\%$ CI, $[l,u]$, and true value, $x$, is calculated as
\[S_\alpha(l,u;x) =(u-l)+\frac{2}{\alpha}(l-x)\mathbbm{1}\{x<l\}+\frac{2}{\alpha}(x-u)\mathbbm{1}\{x>u\},\]
where $\mathbbm{1}$ denotes the indicator function. 
Note that a lower interval score is more desirable. The interval score rewards narrow intervals, but also penalizes intervals that do not contain the truth, striking a balance similar to MSE.
Results for the Gaussian response simulation, including MSE, absolute bias, interval score, and 95\% credible interval (CI) coverage are presented in Table~\ref{tab:gauss_results}.  
All metrics are calculated as averages across areas, weeks, and all 100 sub-samples.

The G-TULM outperforms both the G-BULM and direct estimator in terms of MSE and interval score. 
We note that the direct estimator resulted in the lowest bias, which is to be expected, since the direct estimator is asymptotically design unbiased. 
Regardless, the MSE of the direct estimator is drastically higher than that of either model. 

\begin{figure}[h]
\begin{center}
\includegraphics[width=\linewidth]{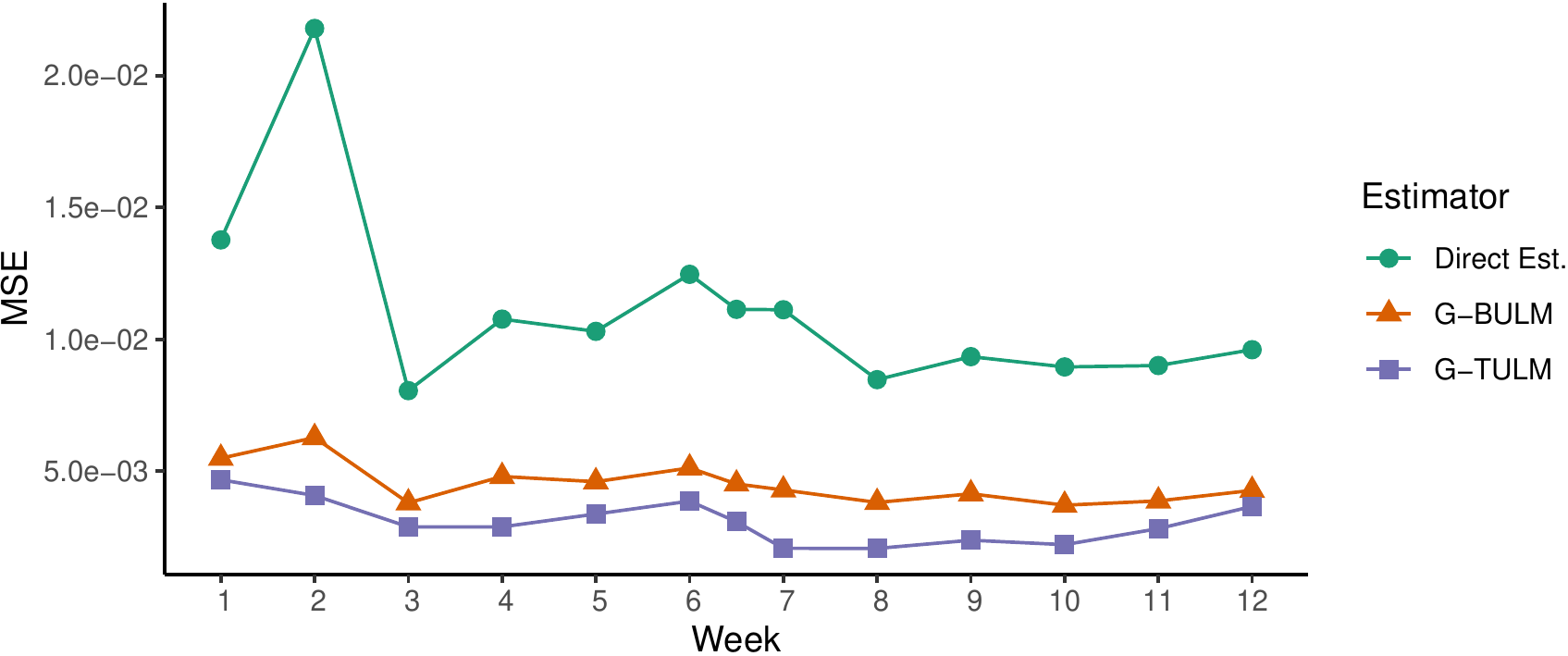}
\end{center}
\caption{Comparison of MSE by week for the Gaussian response models and direct estimators fit to empirically simulated data. For every sub-sample, MSE values are calculated for every area at each week, then averaged across areas and sub-samples.\label{fig:sim_gauss_MSE_time_series}}
\end{figure}

\begin{figure}[h]
\begin{center}
\includegraphics[width=\linewidth]{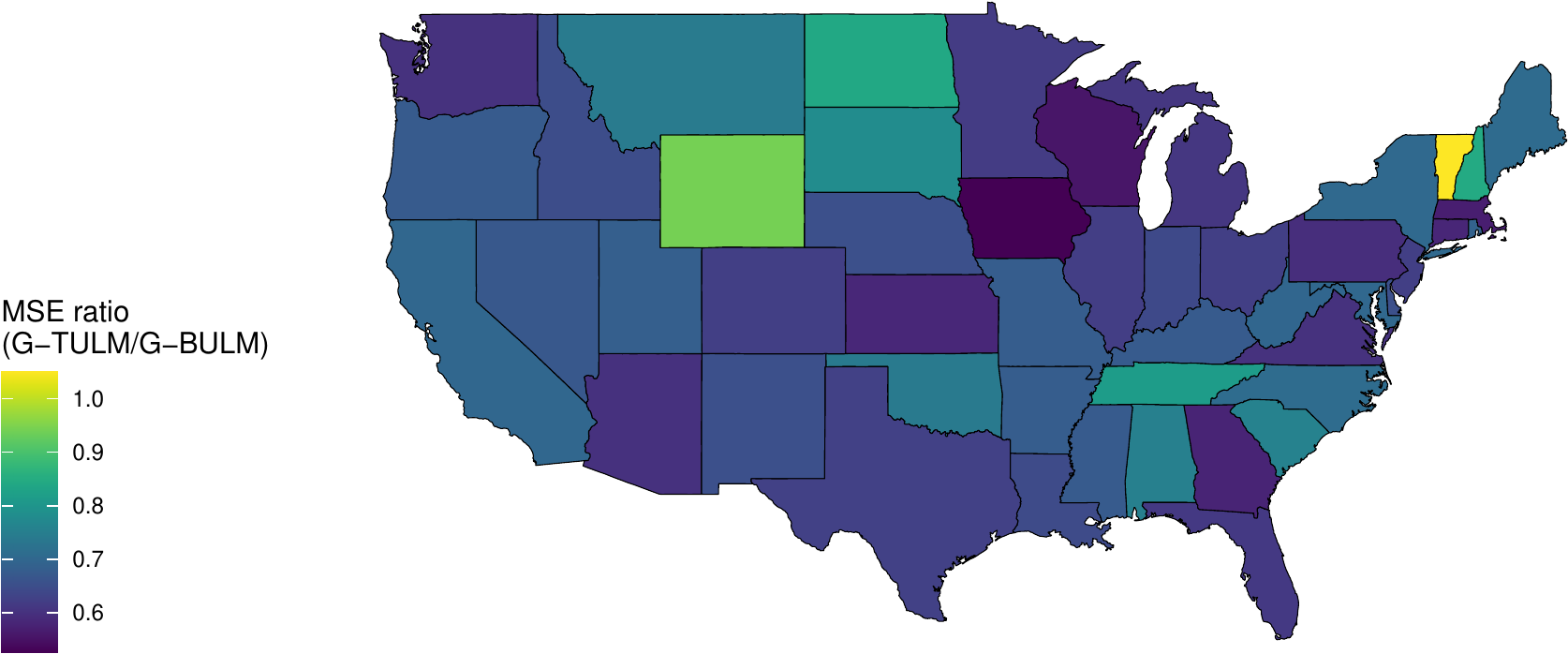}
\end{center}
\caption{Ratio of the G-TULM model MSE to the G-BULM model MSE as fit to the empirically simulated data. 
For every sub-sample, the MSE ratios are calculated for every area at each week, then averaged across weeks and sub-samples. \label{fig:sim_gauss_MSE_ratio}}
\end{figure}

Figure~\ref{fig:sim_gauss_MSE_time_series} %
shows MSE (averaged across states) for both models and the direct estimator at each time point. 
Both models provide a drastic improvement in MSE over the direct estimator at each week. 
To get a better sense of the decrease in MSE that the G-TULM provides relative to the G-BULM, we also plot the ratio of the two model MSEs for all areas averaged across time in Figure~\ref{fig:sim_gauss_MSE_ratio}.
This ratio is less than one for all but a single state, indicating a nearly uniform reduction in MSE for our proposed approach over the baseline model. %

\begin{table}
\caption{Empirical simulation study results for direct- and model-based estimates of the Gaussian response. Values are computed for each week and area then averaged over time, area, and across 100 simulations. CIs and interval scores are calculated for $\alpha=.05.$\label{tab:gauss_results}}
\begin{center}
\begin{tabular}{lrrrr}
  \hline
Method & MSE & Absolute Bias & CI Coverage & Interval Score \\
  \hline
  Direct & $11.14 \times 10^{-3}$      & $8.921 \times 10^{-3}$ & $89\%$      & $5.280 \times 10^{-1}$ \\
  G-BULM & $4.518 \times 10^{-3}$      & $1.379 \times 10^{-2}$      & $96\%$      & $3.352 \times 10^{-1}$ \\
  G-TULM & $\bm{3.086 \times 10^{-3}}$ & $1.759 \times 10^{-2}$      & $95\%$ & $\bm{2.628 \times 10^{-1}}$ \\
   \hline
\end{tabular}
\end{center}
\end{table}

\begin{figure}[h]
\begin{center}
\includegraphics[width=\linewidth]{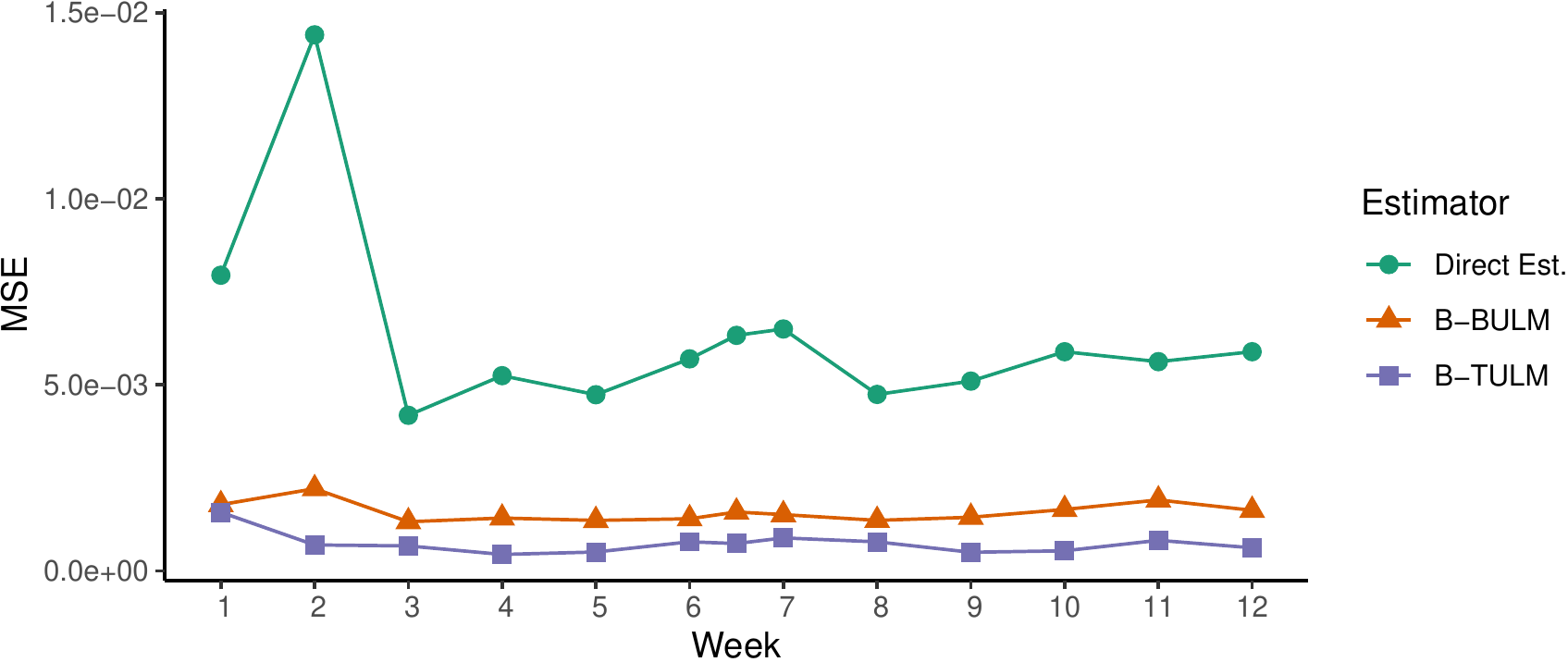}
\end{center}
\caption{Comparison of MSE by week for the binary response models and direct estimators fit to empirically simulated data. For every sub-sample, MSE values are calculated for every area at each week, then averaged across areas and sub-samples.\label{fig:sim_bin_MSE_time_series}}
\end{figure}

\begin{figure}[h]
\begin{center}
\includegraphics[width=\linewidth]{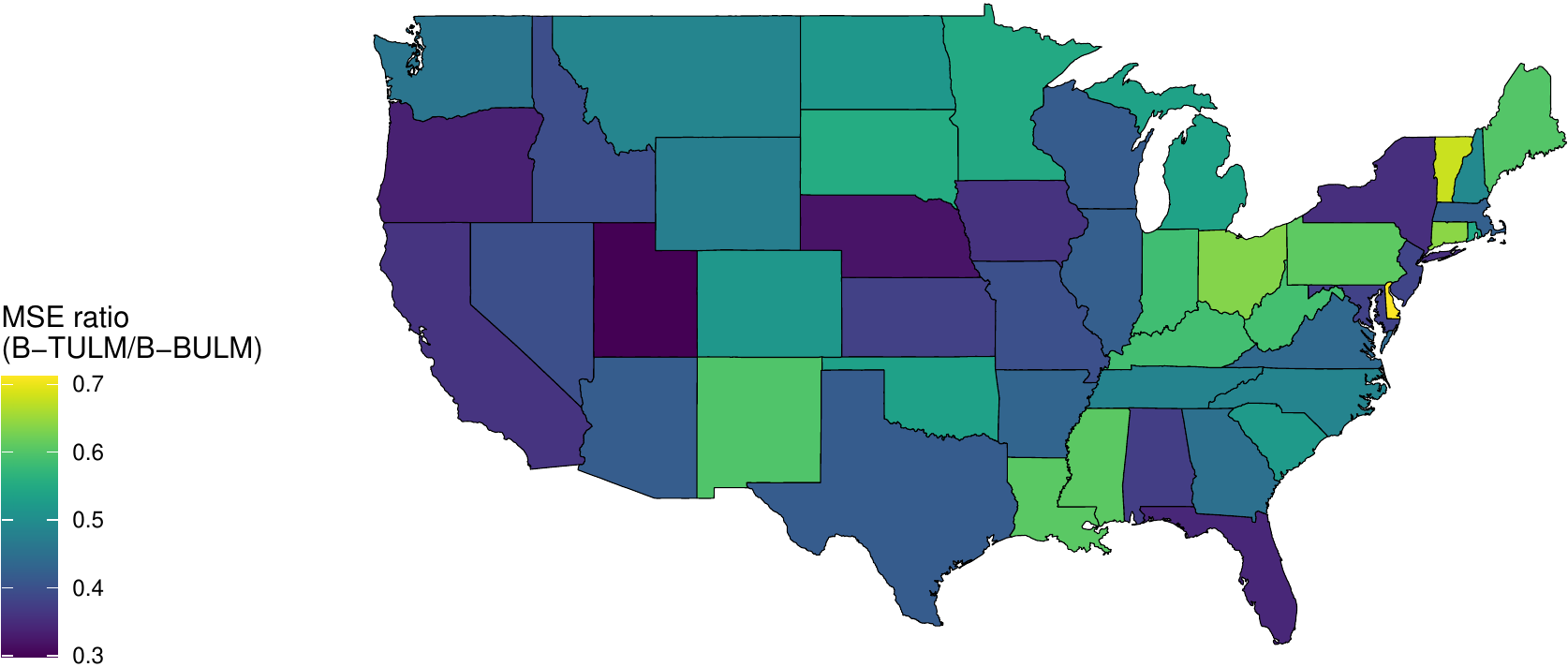}
\end{center}
\caption{Ratio of the B-TULM model MSE to the B-BULM model MSE as fit to the empirically simulated data. 
For every sub-sample, the MSE ratios are calculated for every area at each week, then averaged across weeks and sub-samples. \label{fig:sim_bin_MSE_ratio}}
\end{figure}

Results for the binary response are presented in Table~\ref{tab:binary_results}. Figures~\ref{fig:sim_bin_MSE_time_series} and \ref{fig:sim_bin_MSE_ratio} present an area-level plot and a time series plot of MSEs, respectively. 
In this simulation, the improvement in MSE of the B-TULM over B-BULM is even more dramatic and the proposed B-TULM offers improvement over every time point and across all areas. 
Although the direct estimator CI coverage is closer to the nominal level than for the two models,  the intervals are overly wide, resulting in large interval scores.
Both model-based approaches have considerably lower interval scores than the direct estimator, indicating better predictive coverage. Furthermore, the B-TULM offers a dramatic reduction in both MSE and interval score relative to the B-BULM.

\begin{table}[h]
\caption{Empirical simulation study results for direct- and model-based estimates of the Gaussian response. All metrics are computed for each week and area pair for every sub-sample. These values are then averaged over time, area, and sub-samples. CIs and interval scores are calculated for $\alpha=.05.$\label{tab:binary_results}}
\begin{center}
\begin{tabular}{lrrrr}
  \hline
Method & MSE & Absolute Bias & CI Coverage & Interval Score \\
  \hline
Direct & $6.332 \times 10^{-3}$ & $6.530 \times 10^{-3}$ & $96\%$ & $4.127 \times 10^{-1}$ \\
  B-BULM & $1.586 \times 10^{-3}$ & $1.865 \times 10^{-2}$ & $99\%$ & $2.103 \times 10^{-1}$ \\
  B-TULM & $\bm{7.397 \times 10^{-4}}$ & $1.588 \times 10^{-2}$ & $98\%$ & $\bm{1.324 \times 10^{-1}}$\\ \hline
\end{tabular}
\end{center}
\end{table}

\section{Household Pulse Survey Analysis}\label{sec:hps_analysis}
In this section, we utilize the B-TULM model for our motivating (binary) analysis of ``expected job loss" during the first phase of COVID-19 using the full HPS dataset. 
The simulation in the preceding section shows the benefits of model-based estimation in the case where domain specific sample sizes are low. 
Even in a setting where sample sizes are substantially larger, it is common to want estimates for more granular tabulations. 
In this case, model-based estimates provide a great advantage over design-based estimates (direct estimators), which will have extremely high standard errors for areas with low sample sizes or will be impossible to calculate for areas with no sampled values. 

Again, herein, we consider all responses from the continental U.S.\ for the 12 weeks of phase one of the HPS and analyze ``expected job loss". This variable was the same variable used in our empirical simulation study (Section~\ref{sec:empsim}) and, thus, we use the same set of covariates. After dropping $7,000$ responses due to item non-response, we are left with 1,065,430 responses and 844,547 unique respondents.%

We split age into 13 categories: one category for respondents between 18 to 25 years old, 11 categories of five-year intervals ranging from 26 to 80 years old, and a final category consisting of ages over 81 years old. 
With two categories for sex, 12 weeks, and 49 areas, we require estimates for 15,288 domains (cross-classification cells) in the population. 
At this level of granularity, some  domains are represented by only a single response in the sample. 
This scenario commonly arises for finer levels of geography as well. 
For example, considering a survey which tabulates at the census tract level, it is likely that some tracts may contain only a few responses.

To generate population level estimates, we obtain population size estimates for each cell from the Current Population Survey and generate synthetic response values for each individual in the population via the posterior predictive distribution. 
Since all members of a cell will utilize identical covariate and random effect values, we need only generate posterior predictions for the count of ``yes'' responses for each cell.
In order to account for the within-respondent dependence during post-stratification, we treat each individual in the population as though they were ``unsampled'' prior to week one. 
In all following weeks, we split the cell counts into those who responded ``yes'' and those who responded ``no" in the previous week since these groups take on different synthetic covariate values.  

\begin{figure}[h]
\begin{center}
\includegraphics[width=\linewidth]{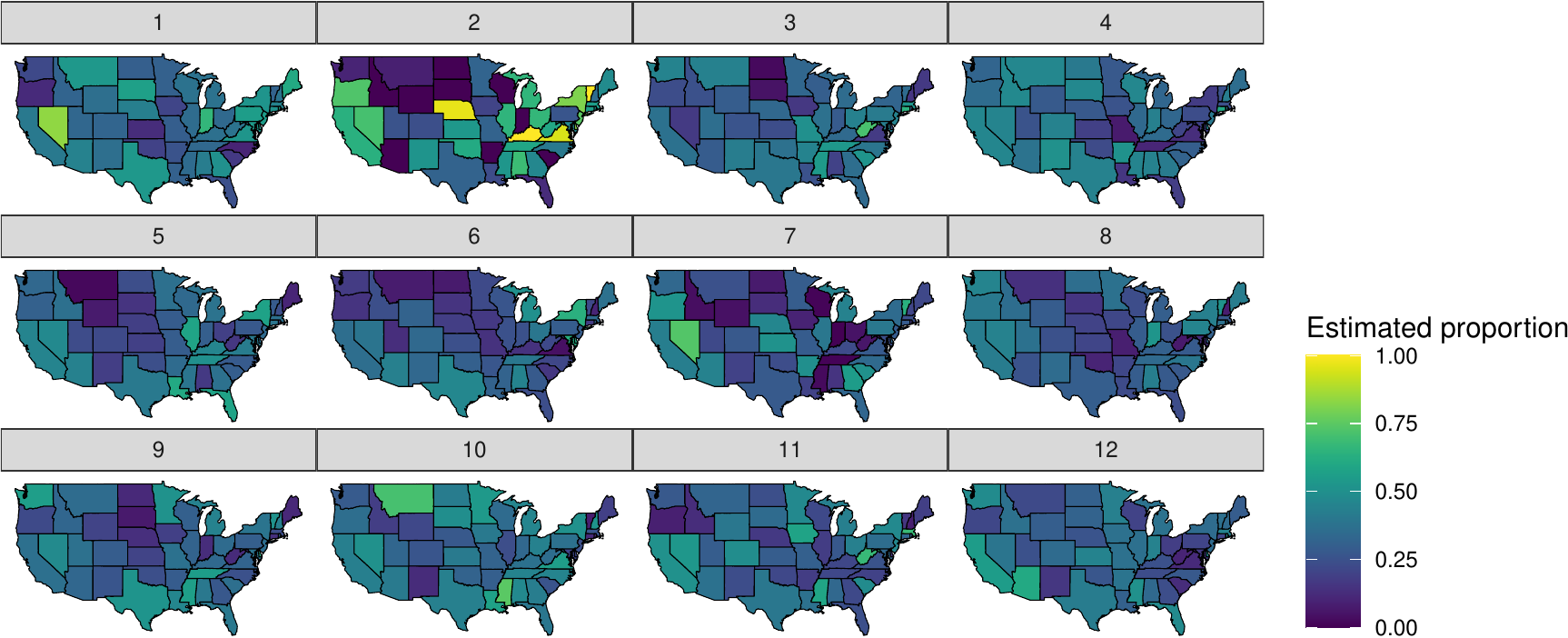}
\end{center}
\caption{Areal time series plot of the direct estimates for proportion of males aged 17 to 25, who expect household job loss, where estimates are calculated from the entire HPS Phase 1 data.
The plot is faceted by week, with the first time point at top-left and the last time-point at bottom-right.\label{fig:data_analysis_dir_est_prop}}
\end{figure}

\begin{figure}[h]
\begin{center}
\includegraphics[width=\linewidth]{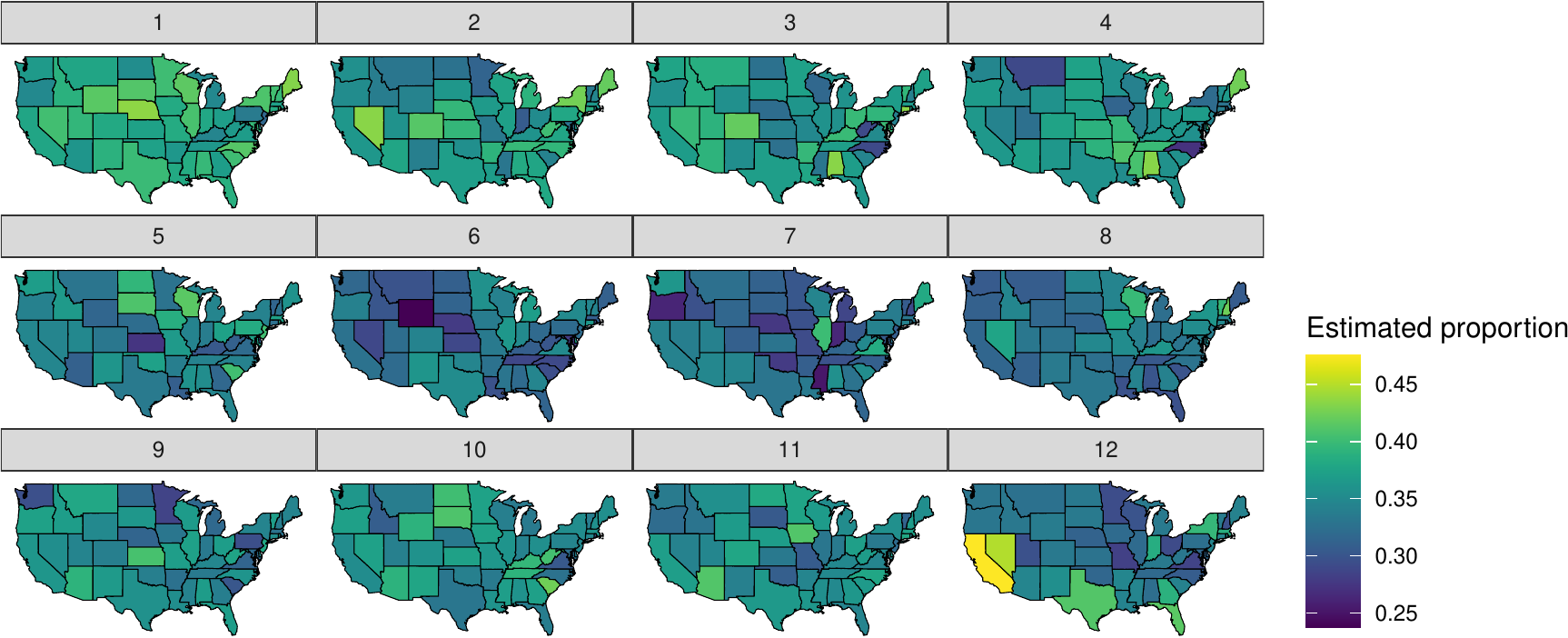}
\end{center}
\caption{Areal time series plot of the B-TULM estimate of proportion of males aged 17 to 25, who expect household job loss, where the model is fit to the entire HPS Phase 1 data. 
The plot is faceted by week, with the first time point at top-left and the last time-point at bottom-right Note the different color scale from Figure~\ref{fig:data_analysis_dir_est_prop}. \label{fig:data_analysis_TULM_prop}}
\end{figure}

We also calculate direct estimates, which we plot for all weeks and areas in Figure~\ref{fig:data_analysis_dir_est_prop}. Figure~\ref{fig:data_analysis_TULM_prop} shows an identical plot for the B-TULM estimates. 
The two plots are on differing scales because the direct estimator produces many estimates near either zero or one, due to low sample sizes. In contrast, the model-based estimates fall in a much narrower range. 
Note that there is a  considerable noise present in the direct estimates. 
For example, the estimated proportion of men aged 17 to 25 in Nebraska, who expect household job loss in week two of the survey is nearly one, while neighboring states are estimated at nearly zero. 
In the following week, states with a high estimated proportion drop drastically. 
Such sharp swings across nearby (presumably similar) areas or short timescales are unrealistic.
Contrast this behavior with that of the B-TULM, model-based estimates, which are much more smooth over space and time and, in general, do not exceed 0.5.

\begin{figure}[h]
\begin{center}
\includegraphics[width=\linewidth]{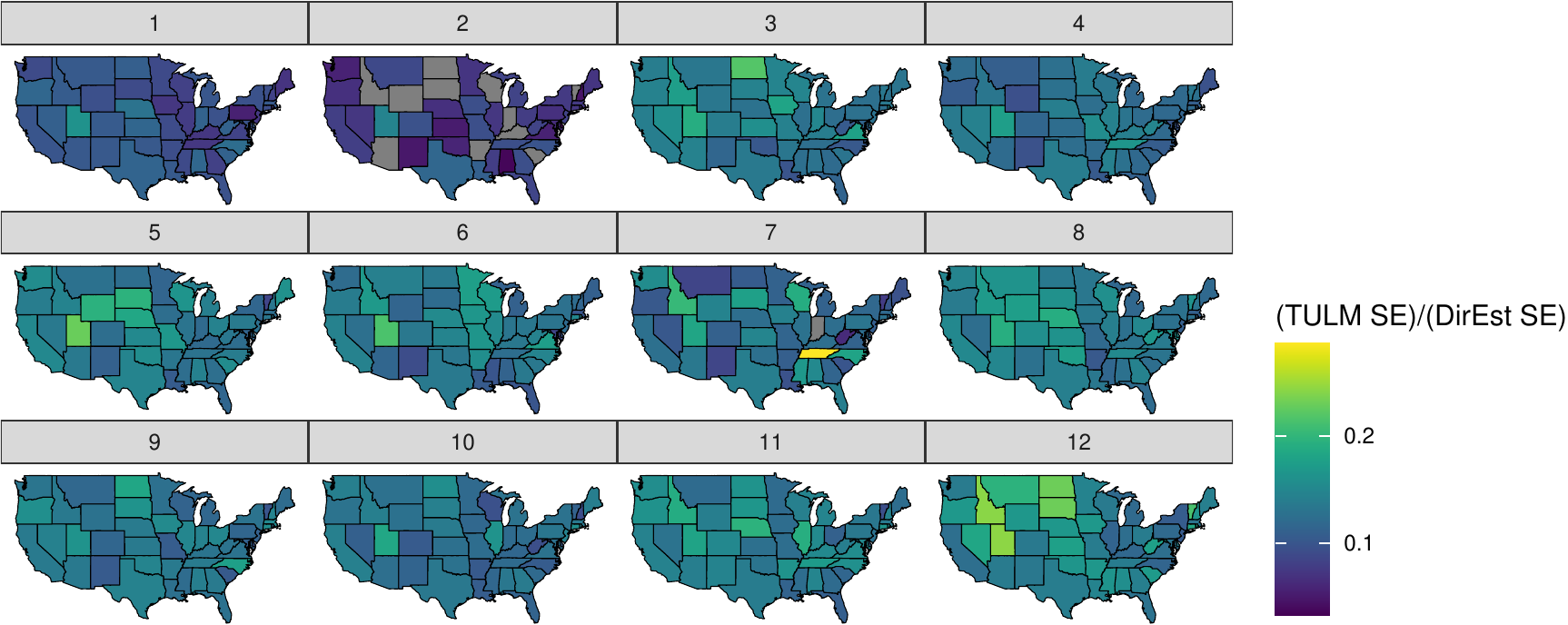}
\end{center}
\caption{Area-level time series plot of the ratio of B-TULM SE to direct estimate SE as fit to the entire HPS Phase 1 data to estimate proportion of males aged 17 to 25, who expect household job loss. 
The plot is faceted by week, with the first time point at top-left and the last time-point at bottom-right. 
Grey areas are those where the sample did not contain enough responses to calculate a direct estimate.
\label{fig:data_analysis_SE_ratio}}
\end{figure}

Figure~\ref{fig:data_analysis_SE_ratio} plots the ratio of B-TULM standard errors to the direct estimate standard errors. This ratio does not exceed $0.3$ for any area or time point, indicating a substantial reduction in the standard errors that is uniform over time and across all areas.

Further examination of the results of our analysis reveals several other interesting findings. 
California, Florida, Georgia, Indiana, Nevada, and Texas all exhibited high expected job loss in week 12.
Recall that data collection for week 12 of the HPS occurred between July 16 and July 21, 2020. 
Hence, data collection for week 11 ended on July 14. Meanwhile, on July 13, a significant shutdown was instituted in California.\footnote{\baselineskip = 10pt \url{https://www.reuters.com/article/us-health-coronavirus-usa/california-shuts-down-businesses-schools-as-coronavirus-outbreak-grows-idUSKCN24E23E}} 
This shutdown could be a contributor to the increase in expected job loss in the state observed in week 12. 
In addition, this date is roughly one week after the Fourth of July holiday weekend and coincides with the date we would expect to see infections develop into cases and hospitalizations.
Furthermore, Florida experienced a single-day death record from COVID-19 on July 26, whereas Texas saw a record for deaths earlier that week. %

The situation in Nevada exhibits another notable pattern. 
In particular, there appears to be a decrease in expected job loss during weeks five and six, which roughly coincides with casino hotels reopening following the initial wave of imposed restrictions.\footnote{\baselineskip=10pt \url{https://www.rgj.com/story/news/2020/07/10/nevada-bars-closed-las-vegas-casinos-open-governor-sisolak-update/5414497002/}} 

\begin{figure}[h]
\begin{center}
\includegraphics[width=\linewidth]{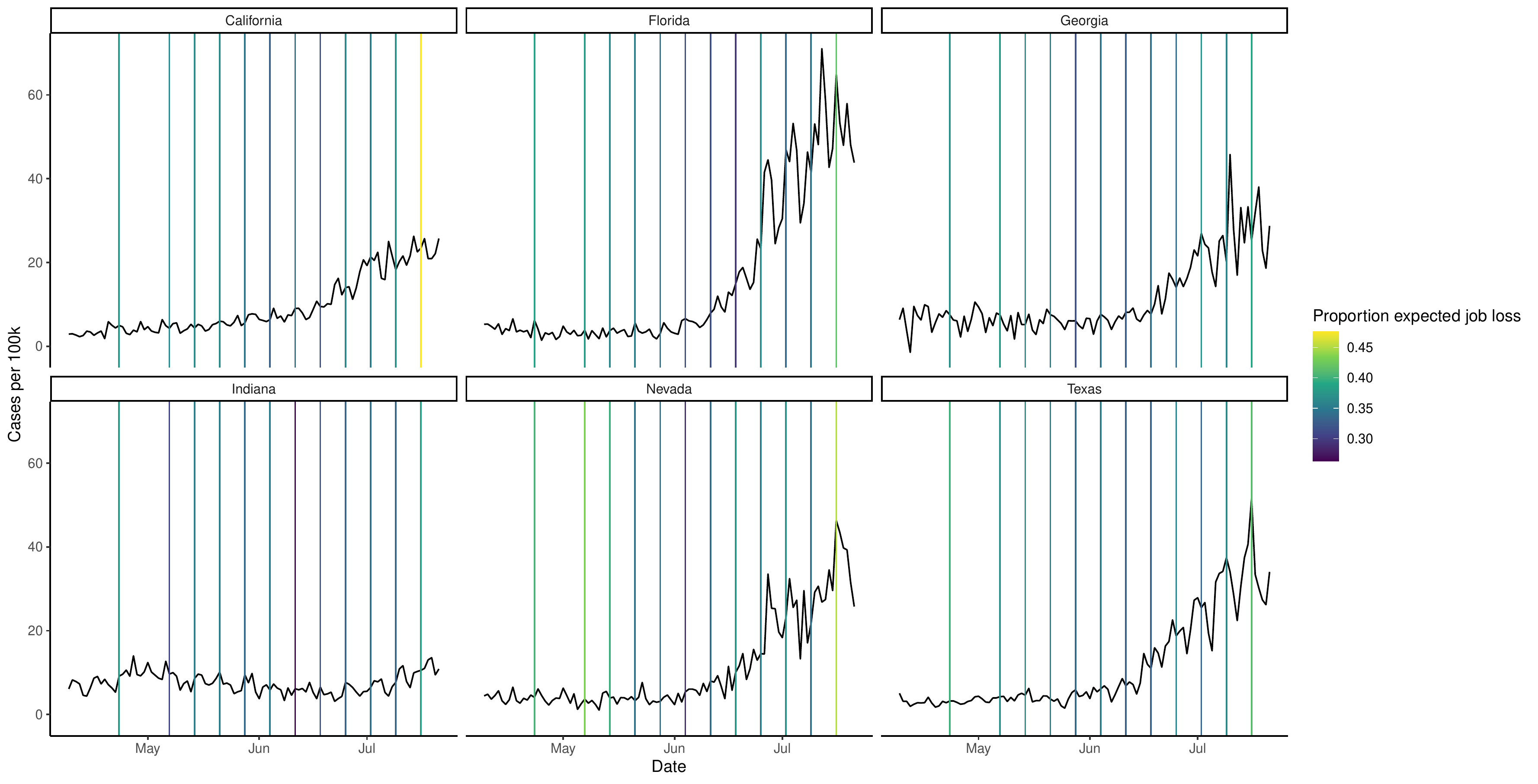}
\end{center}
\caption{COVID-19 cases (per 100K) during HPS Phase 1 for selected states with TULM estimates of expected job loss. Case count data were obtained from the \textit{New York Times} via the covdata R library \citep{covdata}.} \label{fig:covid_cases}
\end{figure}

To better illustrate these trends, Figure~\ref{fig:covid_cases} shows the number of cases (per 100K) by state with TULM estimates of expected job loss superimposed. The vertical bars placed at the end of each week of HPS data collection are shaded to reflect the TULM estimate of the proportion of expected job loss in that given week.

\section{Discussion}
\label{sec:discussion}

This article proposes new Bayesian unit-level methodology for survey data with a longitudinal structure. Specifically, we introduce a multi-scale model for Gaussian responses that accounts for correlation within a units set of responses over time, as well as across geographic domains over time. In addition to this, we introduce a similar model that may be used for binomial data, as many survey variables tend to be binary at the unit-level. In both cases, we emphasize computational efficiency by constructing Gibbs samplers with conditional conjugacy. Importantly, although our simulation studies were conducted on a high-performance computing cluster to shorten runtimes, the models proposed herein are computationally efficient and can be fit on a standalone laptop or workstation with moderate computing resources.

Through simulation studies, we show that our proposed approach can yield more precise point and interval estimates when compared to either traditional direct estimators,  or existing model-based estimators that lack a longitudinal or temporal component. In addition to this, we use the proposed methodology in order to construct estimates of expected job loss during the early weeks of the COVID-19 pandemic, using data from the HPS. The constructed estimates offer greater precision than direct estimates, and could be valuable in assessing the societal and economic impact of the pandemic.

 We use data from Phase 1 of the HPS, however, the design and structure of the HPS has been revised since that time. For example, starting in Phase 2, respondents were not re-selected in subsequent weeks. Thus, some modifications to our proposed methodology may be necessary in order to construct estimates using data beyond Phase 1. Nevertheless, joint analysis of Phase 1 data with data beyond Phase 1, could benefit from the proposed methodology. 

 Our construction of models for both Gaussian and Binomial data allows for estimation using a wide variety of responses from the HPS. Still, there are other response types that do not immediately fit into our current framework. For example, higher order categorical variables are common in many surveys. Thus, extending the methodology to multinomial or ordinal response types is a subject of future work.

 Finally, although this work was heavily motivated by the HPS, other surveys exist with longitudinal structure, such as the Survey of Income and Program Participation (SIPP), the Current Population Survey (CPS), and the Monthly Retail Trade Survey, among others. In some cases, these surveys utilize more complex panel structures than what was used in Phase 1 of the HPS. Thus, this work is of independent interest, and it is likely that our methodology could prove valuable for producing estimates with data from other surveys, perhaps with some slight modifications. Analysis of data from alternative surveys with longitudinal structure is another avenue of future work.

\if0\blind
{
  \section*{Acknowledgements}
  This article is released to inform interested parties of ongoing research and to encourage discussion. The views expressed on statistical issues are those of the authors and not those of the NSF or U.S. Census Bureau.
  

  \section*{Funding}
   This research was partially supported by the U.S. National Science Foundation (NSF) under NSF grants SES-1853096, NCSE-2215168, and NCSE-2215169.
} \fi

\clearpage

\bibliography{writeup}
\bibliographystyle{jasa}

\appendix
\section{Appendix}
\label{sec:appendix}
\subsection{Sampling algorithm for the G-TULM}
\subsubsection{Notation}
Describing the full conditionals requires some additional notation, which we introduce here.
We define $\bm W$ to be the diagonal matrix of survey weights for all respondents across all time points with $(\bm W)_{ii}=w_{ii}$,
\[\bm W = (w_{11}, w_{21},\ldots, w_{n_11}, w_{12}, w_{22}\ldots, w_{n_TT}) \]
Similarly, let $\bm \Psi$ be the $n\times m$ matrix of incidence vectors with $(\bm\Psi)_{ij}=1$ if the unit corresponding to row $i$ belongs to area $j$ and all other entries set to zero.
Let $\bm X$ be the design matrix whose $i$th row is the $p-$dimensional vector of covariates $\bm x_i$.
Lastly, we take $\bm \eta$ to be the concatenation of the vector of random effects for each time,
\[\bm \eta = (\bm \eta_1', \bm \eta_2',\ldots, \bm \eta_T') = (\eta_{11}, \eta_{21},\ldots, \eta_{n_11}, \eta_{12}, \eta_{22}\ldots, \eta_{n_TT}).\]

For each of these matrices and vectors, we denote with a subscript $t$ the subset of entries corresponding to responses collected at time $t$.
We denote with a superscript $f$ the subset of entries corresponding to first-time respondents and with a superscript $r$ the subset corresponding to follow-up respondents.
A number of combinations of sub- and superscripts are possible.
For example,
\begin{itemize}
\item $\bm X_t^r$ is the matrix of covariates for all follow-up respondents sampled at time $t$.
\item $\bm W^f$ is the matrix of survey weights corresponding to all follow-up respondents in the sample across all times
\item $\bm \Psi_2$ is the matrix of incidence vectors for all respondents (both first-time and follow-up) at time 2. 
\item $\bm \eta$ refers to the random effects vector for all responses across all times.
\end{itemize}

For the vector of responses $\bm y$, we define the additional superscripts $r_1$ and $r_2$, such that
$\bm y^{r_2}$ is the vector of all follow-up responses and $\bm y^{r_1}$ is a vector of the corresponding responses from previous weeks.
Under this notation, the first response from a household $i$ is recorded as entry $y_i^f$.
If the household provides a second response, that response is denoted $y_{i'}^{r_2}$ and the first response is reproduced as $y_{i'}^{r_1}$, with $i'$ not necessarily equal to $i$. 
A third response will be recorded as another entry $y_{i''}^{r_2}$ and $y_{i''}^{r_1}=y_{i'}^{r_2}$ will reproduce the second response. 
This ``duplication'' simplifies indexing when we need to reference a follow-up response and the corresponding response from that unit in the previous week.

For the binary case we also have  the vector of latent variables for the P\'olya-Gamma data augmentation scheme
\[\bm\Omega = \text{diag}(\omega_{11}, \omega_{21}, \ldots, \omega_{1T}, \omega_{2T},\ldots,\omega_{n_TT})\] and the additional parameter vector
\[\bm \kappa=(\widetilde{w}_{11}(y_{11}-1/2), \widetilde{w}_{12}(y_{12}-1/2), \ldots, \widetilde{w}_{1T}(y_{1T}-1/2), \widetilde{w}_{21}(y_{21}-1/2),\ldots,\widetilde{w}_{n_TT}(y_{n_TT}-1/2))'.\]

As before, for any given $t$, let $\bm\Omega_t = \text{diag}(\omega_{1t},\omega_{2t},\ldots,\omega_{{n_t}t})$; that is, the entries of $\bm\Omega$ corresponding only to time $t$.
Likewise, 
let $\bm\kappa_t = (\kappa_{1t},\kappa_{2t},\ldots,\kappa_{n_tt})$.
Lastly, note that $\bm\kappa/\bm\omega$ represents element-wise division. 

\subsubsection{Algorithm}
\label{sec:append_gauss}

\begin{enumerate}
\item Sample $\bm \beta$:
  \[\bm \beta|\cdot \sim  \begin{aligned}[t] N_p\Bigg(\bm\mu_\beta &=\sigma_\beta\left(\frac{1}{\sigma^2}(X^f)'W^f\bm u_\beta -   \frac{1}{\sigma^2(1+\rho)}(X^r)'W^r\bm v_\beta\right),\\
                                 \Sigma_\beta &=\left(\frac{1}{\sigma^2_\beta}I_p + 
                                                    \frac{1}{\sigma^2}(X^f)'W^fX^f -                                                    \frac{1-\rho}{\sigma^2(1+\rho)}(X^r)'W^rX^r                             \right)^{-1} \Bigg)\end{aligned},\]
    where $\bm u_\beta = \bm y^f-\bm\eta^f$ and $\bm v_\beta = \bm y^{r_2}-\bm\eta^{r_2} - \rho(\bm y^{r_1} -\bm\eta^{r_1} ).$
\item Sample $\bm \eta_1$:
 \[\bm \eta_1|\cdot \sim N_m
        \begin{aligned}[t]
           \Bigg( 
                     \bm\mu_{\eta_1} &= \Sigma_{\eta_1} \left(\frac{1}{\sigma^2}(\Psi_1^f)'W_1^f\bm u_{\eta_1} + \frac{\rho}{\sigma^2(1-\rho^2)}(\Psi_2^r)'W_2^r\bm v_{\eta_1}  + \frac{\phi}{\sigma^2_\eta}\bm\eta_2\right),\\ 
                     \Sigma_{\eta_1} &= \left( \frac{1}{\sigma^2}(\Psi_1^f)'W_1^f\Psi_1^f + \frac{\rho^2}{\sigma^2(1-\rho^2)}\Psi_2^rW_2^r\Psi_2^r  + \left(\frac{1}{100}+\frac{\phi^2}{\sigma^2_\eta}\right)I_m \right)^{-1}
           \Bigg)
        \end{aligned}\]
    where $\bm u_{\eta_1} = y^f - X^f\bm\beta$ and $\bm v_{\eta_t} = -(\bm y_2^{r_2}-X_2^r\bm \beta-\Psi_2^r\bm\eta_2 - \rho(\bm y_2^{r_1}-X_2^r\bm \beta )).$
\item Sample $\bm\eta_t$ for $t=2,\ldots, T-1$:
     \[\begin{aligned}[t]
      \bm \eta_t|\cdot \sim N_m
           \Bigg(\bm\mu_{\eta_t} &= \Sigma_{\eta_t}\bigg(\frac{\phi}{\sigma^2_\eta}(\bm\eta_{t-1}+\bm\eta_{t+1})\\
                                                    &    + \frac{1}{\sigma^2}\left((\Psi_t^f)'W_t^f\bm u_{\eta_t} 
                                                         + \frac{1}{1-\rho^2}\left((\Psi_t^r)'W_t^r\bm v_{\eta_t}
                                                         -\rho (\Psi_{t+1}^r)'W_{t+1}^r\bm w_{\eta_t} \right)\right)\bigg),\\ 
                 \Sigma_{\eta_t} &= \bigg(\frac{1+\phi^2}{\sigma^2_\eta}I_m\\
                                                  &+ \frac{1}{\sigma^2}\left((\Psi_t^f)'W_t^f\Psi_t^f + \frac{1}{1-\rho^2}\left(\left(\Psi_t^r)'W_t^r\Psi_t^r+\rho^2(\Psi_{t+1}^r)'W_{t+1}^r\Psi_{t+1}^r\right)\right)\right)\bigg)^{-1}
            \Bigg),
     \end{aligned},\]
     where \begin{itemize}
             \item $\bm u_{\eta_t} = \bm y_t^f-X_t^f\bm\beta,$
              \item $\bm v_{\eta_t}= \bm y_t^{r_2}-(1-\rho)X_t^r\bm\beta - \rho(\bm y_t^{r_1}-\Psi_t^r\bm\eta_{t-1}),$
            \item $\bm w_{\eta_t}=\Psi_{t+1}^r\bm\eta_{t+1} + (1-\rho)X^{r_2}\bm\beta - (\bm y_{t+1}^{r_2}-\rho\bm y_{t+1}^{r_1}).$
            \end{itemize}
\item Sample $\bm \eta_T$:
      \[\begin{aligned}[t]
        \bm \eta_T|\cdot \sim N_m
        \Bigg(\bm\mu_{\eta_T} &=  \Sigma_{\eta_T} \left(\frac{\phi}{\sigma^2_\eta}\bm\eta_{T-1} 
        + \frac{1}{\sigma^2}\left((\Psi_T^f)W_T^f\bm u_{\eta_T} 
        + \frac{1}{1-\rho^2}(\Psi_T^r)W_T^r\bm v_{\eta_T}\right)\right)\\
              \Sigma_{\eta_T} &=  \left(\frac{1}{\sigma^2_\eta}I_m + \frac{1}{\sigma^2}\left((\Psi_T^f)'W_T^f\Psi_T^f + \frac{1}{1-\rho^2}((\Psi_T^{r})'W_T^r\Psi_T^{r})\right)\right)^{-1}
        \Bigg)
       \end{aligned},\]
       where $\bm u_{\eta_T}=\bm y_T^f -X_T^f\bm\beta$ and $\bm v_{\eta_T} = \bm y_T^{r_2} - X_T^{r_2}\bm\beta - \rho(\bm y_T^{r_1}-X_T^r\bm\beta -\Psi_T^r\bm\eta_{T-1}).$
\item Sample $\rho$ with a Metropolis step. The full conditional is 
 \begin{align*}
p(\rho|\cdot)&\propto (1-\rho^2)^{-N_r/2}\\
                      &\times \exp\left\{-\frac{1}{2\sigma^2(1-\rho^2)}
                                 \left((\rho^2-\rho)\tilde{\bm y}^{r_1}-\rho\tilde{\bm y}^{r_2})'
                                     W^r((\rho^2-\rho)\tilde{\bm y}^{r_1}-\rho\tilde{\bm y}^{r_2})\right)\right\}\\
                      &\times \mathbbm{I}(-1<\rho<1).
 \end{align*}
 This cannot be sampled from directly so we use a random-walk Metropolis algorithm with a uniform proposal distribution centered on the previous draw of $\rho.$
\item Sample $\phi$: 
 \[ \phi |\cdot \sim \text{TruncNorm}\left(\frac{\sum_{t=2}^T\bm\eta_t'\bm\eta_{t-1}}{\sum_{t=2}^T\bm\eta_{t-1}'\bm\eta_{t-1}},\frac{\sigma^2_\eta}{\sum_{t=2}^T\bm\eta_{t-1}'\bm\eta_{t-1}},-1,1\right)\]
 \item Sample $\sigma^2$: \[\sigma^2|\cdot \sim \text{IG}\left(1+\frac{N}{2}, 1+\frac{1}{2} (\widetilde{\bm y}^f)' W^f\widetilde{\bm y}^f + \frac{1}{2(1-\rho^2)} (\widetilde{\bm y}^{r_2} - \rho\widetilde{\bm y}^{r_1})'W^r(\widetilde{\bm y}^{r_2} - \rho\widetilde{\bm y}^{r_1})\right)\]
\item Sample $\sigma^2_{\eta_1}$:
     \[\sigma^2_{\eta_1} | \cdot \sim \text{IG}\left(a+\frac{m(T-1)}{2}, b + \frac{1}{2}\bm\eta_1'\bm\eta_1\right)\]
\item Sample $\sigma^2_{\eta}$:
    \[\sigma^2_\eta | \cdot \sim \text{IG}\left(a +\frac{m(T-1)}{2}, b + \frac{1}{2}\sum_{t=2}^T(\bm \eta_t-\phi\bm\eta_{t-1})'(\bm \eta_t-\phi\bm\eta_{t-1})\right)\]
\end{enumerate}

\subsection{Sampling algorithm for the B-TULM}
\label{sec:append_binary}

\begin{enumerate}
\item Sample $\omega_{it}$ for $t=1,\ldots,T$, for $i=1,\ldots,n_t$: 
\[\omega_{it}|\cdot \sim \text{PG}(\widetilde{w}_{it}  n_{it},\bm x_i/\bm\beta +\bm\psi_i'\bm\eta_t).\]
\item Sample $\bm \eta_1$:
    \[\bm\eta_1|\cdot \sim \text N_m \left(\bm\mu =\bm\Sigma\left( \bm\Psi_1'\bm\Omega_1(\bm\kappa_1/\bm\omega_1-X_1\bm\beta) + \frac{\phi}{\sigma^2_\eta}\bm\eta_2\right), 
                                \bm\Sigma =\left(\Psi_1'\bm\Omega_1\Psi_1 + \left(\frac{1}{\sigma^2_{\eta_1}} + \frac{\phi^2}{\sigma^2_\eta}\right) I_m\right)^{-1}\right).\]
\item Sample $\bm\eta_t$ for $t=2,\ldots, T-1$:
\[\bm\eta_t |\cdot \sim \begin{aligned}[t]\text{N}_p\Bigg(\bm\mu &= \bm\Sigma \left( \Psi_t'\bm\Omega_t(\bm\kappa_t/\bm\omega_t - X_t\bm\beta)+\frac{\phi}{\sigma^2_\eta}(\bm\eta_{t-1}+\bm\eta_{t+1})\right),\\
                                         \bm\Sigma&=\left(\Psi_t'\bm\Omega_t\Psi_t + \left(\frac{1}{\sigma^2_{\eta}}+\frac{\phi^2}{\sigma^2_\eta}\right)I_m\right)^{-1}\Bigg).\end{aligned}\]
\item Sample $\eta_T$:
\[\bm\eta_T |\cdot \sim \text{N}_p\left(\bm\mu = \bm\Sigma \left( \Psi_T'\bm\Omega_T(\bm\kappa_T/\bm\omega_T - X_T\bm\beta)+\frac{\phi}{\sigma^2_\eta}\bm\eta_{T-1}\right), 
                                          \bm\Sigma=\left(\Psi_T'\bm\Omega_T\Psi_t + \frac{1}{\sigma^2_\eta}I_m\right)^{-1}\right).\]
\item Sample $\bm\beta$:
\[\bm{\beta}|\cdot \sim \text{N}_p\left(\bm\mu = \bm \Sigma\left(\bm X'\bm\Omega(\bm\kappa/\bm\omega-\bm\Psi^f\bm\eta^f\right), 
                              \bm\Sigma = \left(\bm X'\bm\Omega\bm X + \frac{1}{\sigma^2_\beta}I_p\right)^{-1}  \right).\]
\item Sample $\phi$:
\[\phi|\cdot \sim \text{TruncNorm}\left(\frac{\sum_{t=2}^T\bm\eta_t'\bm\eta_{t-1}}{\sum_{t=2}^T\bm\eta_{t-1}'\bm\eta_{t-1}},
                 \frac{\sigma^2_\eta}{\sum_{t=2}^T\bm\eta_{t-1}'\bm\eta_{t-1}},
                 -1,1 \right).\]
\item Sample $\sigma^2_{\eta_1}$:
        \[\sigma^2_{\eta_1} | \cdot \sim \text{IG}\left(a+\frac{m}{2}, 
                                        b+\frac{\bm\eta_1'\bm\eta_1}{2}\right).\]
\item Sample $\sigma^2_{\eta}$:
       \[\sigma^2_\eta|\cdot \sim \text{IG}\left(a+\frac{m(T-1)}{2}, 
                                  b + \frac{1}{2}\sum_{t=2}^T(\bm\eta_t-\phi\bm\eta_{t-1})'(\bm\eta_t-\phi\bm\eta_{t-1}) \right).\]
\end{enumerate}

\end{document}